\documentclass[12pt]{article}
\input epsf.tex 
\def\beq{\begin{equation}} 
\def\eeq{\end{equation}} 
\def\eeq{\end{equation}} 
\def\bea{\begin{eqnarray}} 
\def\eea{\end{eqnarray}} 
\def\bq{\begin{quote}} 
\def\eq{\end{quote}}

\newcommand{\gsim}{\stackrel{>}{_\sim}}
\newcommand{\lsim}{\stackrel{<}{_\sim}}
\def\ginomsb{\~{g}MSB}
\def\dsl{\not\!\partial}

\makeatletter 
\def\vereq#1#2{\lower3pt\vbox{\baselineskip1.5pt \lineskip1.5pt 
\ialign{$\m@th#1\hfill##\hfil$\crcr#2\crcr\sim\crcr}}} 
\makeatother

\title{Deconstructing Gaugino Mediation}
 
\author{ 
  H.-C. Cheng,$^{a}$\,\thanks{\tt hcheng@theory.uchicago.edu}\ \  
 D.E. Kaplan,$^{a,b}$\,\thanks{\tt dkaplan@theory.uchicago.edu}\ \
  M. Schmaltz,$^c$\,\thanks{\tt schmaltz@fnal.gov}\ \
  W. Skiba$^d$\,\thanks{\tt skiba@mit.edu}\\ \\ \\ 
        \small \sl $^a$\ Enrico Fermi Institute, The University of Chicago,
                         Chicago, IL 60637\\ 
         \small \sl $^b$\ HEP Division, Argonne National
                          Laboratory, Argonne, IL 60439\\ 
       \small \sl $^c$\ Theoretical Physics Department, Fermilab, Batavia,
                        IL 60510\ \\ 
       \small \sl $^d$\ CTP, Massachusetts
                        Institute of Technology, Cambridge, MA 02139\\ \\ 
       } 
 
\begin{document} 
\baselineskip=17pt 
\pagestyle{plain}

\begin{titlepage} 
\vskip-.4in 
\maketitle 
\begin{picture}(0,0)(0,0) 
\put(328,370){EFI-01-24}
\put(268,350){ANL-HEP-PR-01-046}
\put(240,330){FERMILAB-Pub-01/108-T} 
\put(300,310){MIT-CTP-3153} 
\end{picture}

\begin{abstract} 
\leftskip-.6in 
\rightskip-.6in 
\vskip.4in

We present a model of supersymmetry breaking which produces
gaugino masses and negligible scalar masses at a high scale.
The model is inspired by ``deconstructing'' or ``latticizing''
models in extra dimensions where supersymmetry breaking
and visible matter are spatially separated.
We find a simple four-dimensional model which only requires
two lattice sites (or gauge groups)
to reproduce the phenomenology.

\end{abstract} 
\thispagestyle{empty} 
\setcounter{page}{0} 
\end{titlepage}

\section{Introduction} 

Gauge mediated supersymmetry breaking \cite{gmsb} (GMSB) with a messenger scale
much lower than the Planck scale is an elegant solution to the supersymmetric (SUSY) 
flavor problem.  It relies on the fact that standard model (SM) gauge couplings 
are flavor universal and therefore all gauge-loop contributions to soft masses 
produce a degenerate spectrum.

Gaugino mediated supersymmetry breaking (\ginomsb) \cite{ginomsb} is another simple
solution to the SUSY flavor problem.  It uses locality to forbid flavor violating
scalar masses by putting matter of the minimal supersymmetric standard model (MSSM)
on a brane separated in an extra dimension from a brane on which SUSY is broken \cite{amsb}.
Putting gauge superfields in the bulk allows the gauginos to acquire a mass from
direct interactions with the SUSY-breaking brane.  The renormalization group (RG)
is responsible for scalar masses at the low scale and again it is the universality
of gauge couplings which make the generations degenerate.

In this letter, we describe a simple model of ``deconstructed'' gaugino mediation.
In \cite{ACG1,HPW} it was shown that extra dimensional gauge theories can be 
regularized by a lattice of gauge groups connected by ``link'' fields in bifundamental
representations of neighboring groups which break the ``chain'' down to the diagonal
subgroup.  At low energies, these models look identical to extra-dimensional theories
as one finds the correct spectrum of Kaluza-Klein modes.  Using this framework,
intuition gained from extra dimensions can now be applied to constructing interesting
four-dimensional theories \cite{ACG2,cegk}.  In \cite{cegk}, \ginomsb\ was translated 
into this language and shown to reproduce the same spectrum, from a completely
four-dimensional theory. Some of the results of our paper overlap with \cite{cegk}.

We discuss a particularly simple version of deconstructed
gaugino mediation with only two gauge groups in the chain. We examine the
model in the limit of $M_{mess}\gg v$, where $M_{mess}$ is the messenger
scale and $v$ is the scale at which the chain is broken to the diagonal
subgroup.  We find that this allows sufficient separation of SUSY breaking 
and MSSM matter fields to suppress scalar masses as in gaugino mediation.

The model has a striking signature: gaugino masses unify at the GUT scale whereas
all MSSM scalar masses run to zero at an intermediate scale. 

In the next section we describe the model and derive the superpartner spectrum.
In Section 3, we discuss threshold corrections to scalar masses and show they are
negligible compared to that from running.  In Section 4 we show that gauge coupling
unification is preserved in this model.  The final section is reserved for concluding
remarks.

\section{The model}

\begin{figure}[htb]
\vskip 0.0truein
\centerline{\epsfysize=0.8in
{\epsffile{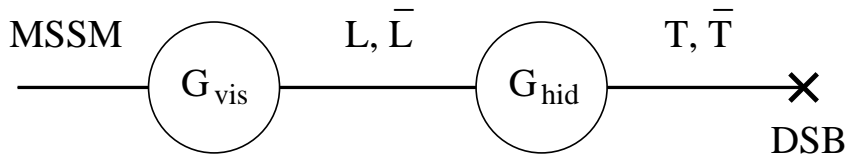}}}
\vskip 0.0truein
\caption[]{\it A schematic diagram of the model. Circles represent gauge groups, lines 
fields charged under these gauge groups. The cross represents the sector where supersymmetry
is dynamically broken. }
\label{fig:loop}
\end{figure}

The model consists of a ``visible'' sector with the standard
model gauge group and the MSSM matter fields. The visible
sector communicates with a ``hidden'' sector gauge group through
``link'' fields which are charged under both.
We assume that SUSY is broken dynamically and SUSY breaking
is mediated to the hidden sector by messenger fields which carry hidden
sector gauge quantum numbers (see Figure 1).

A summary of the dynamics of the model, from the highest scale
to the lowest, is as follows:
SUSY breaking is mediated to the hidden sector gauge group
at the scale $M$, the mass of the messengers. At this scale,
hidden sector gauginos obtain a mass at one loop and the link chiral
superfields pick up masses at two loops. The MSSM fields also obtain
SUSY breaking masses at this scale, but they are suppressed by additional
loop factors and are therefore negligible. At the scale $v$ the link
fields get their vacuum expectation values (vevs) and break the visible
and hidden sector gauge groups to diagonal $SU(3)\times SU(2)\times U(1)$.
The only fields lighter than this scale are the matter and gauge
fields of the MSSM.  

The MSSM gauge fields and gauginos which remain after the symmetry
breaking are linear combinations of the visible and
hidden sector fields. The gauginos inherit a large SUSY breaking
mass from their hidden component at tree level. 
The MSSM scalar masses are suppressed by loop factors compared to the gaugino mass
at this scale, thus reproducing the
pattern of soft masses characteristic of gaugino mediation with compactification scale
$R^{-1} \leftrightarrow v$.

We now present the model in detail. The visible sector gauge
group is $G_{vis}=SU(3)\times SU(2)\times U(1)$ (we discuss
unification in Section 4). MSSM matter fields are charged
under this gauge group with the usual quantum numbers. The hidden sector gauge group
is $SU(5)_{hid}$. Link fields $L+\overline{L}$
transform as $(5,\overline 5)+(\overline 5, 5)$
under $G_{vis}\times SU(5)_{hid}$, where we use $SU(5)$-GUT notation to denote
the representations of $G_{vis}$.
The messenger fields $T+\overline T$ transform as $5+\overline 5$ under
$SU(5)_{hid}$ only. The superpotential is
\beq
\label{superpot}
W=W_{\rm MSSM}+ X T\overline T + S(\frac15 {\rm tr} L\overline L -v^2) +
         {\rm tr} LA\overline L \ ,
\eeq
where $X=M+\theta^2 F$ is the spurion which encapsulates the SUSY breaking
sector and $A$ is an adjoint of $SU(5)_{hid}$.  Meanwhile, $W_{\rm MSSM}$ is the usual 
MSSM superpotential with Yukawa couplings and the $\mu$-term.  
The second term gives the messengers a SUSY violating
mass when $X$ is replaced by its vev.  The third
term forces vevs for the links $L$, $\overline L$ at the scale $v$ and breaks
$G_{vis}\times SU(5)_{hid}$ to the diagonal $SU(3)\times SU(2)\times U(1)$.
One linear combination of the link fields get eaten by gauge fields of the
broken generators and the other gets a mass due to the fourth term.
No light fields charged under the MSSM gauge group remain at scales lower than $v$.

At the messenger scale $M$, the hidden gauginos get their masses at one loop 
as in GMSB: $m_{1/2}^{hid} = (\alpha_{hid}/4\pi) (F/M)$ (we have assumed
only one pair of messengers).  The link fields
$L,\overline{L}$ get scalar masses at two loops.  The
gauginos of $G_{vis}$ are effectively massless at this scale
(gaugino ``screening'' \cite{screening}), and the
leading contribution to scalar masses are four-loop diagrams such as the 
one in Figure 2(a).

Between the scales $M$ and $v$, the soft terms run via the renormalization group.
The visible sector soft masses remain negligible if $M>>v$. In principle,
there could be a one-loop contribution to the MSSM scalar masses from the hypercharge
D-term, however the D-term vanishes at leading order
in our model because the link field scalar masses
are universal.

At the scale $v$, $G_{vis}\times SU(5)_{hid}$ is broken to the diagonal 
$SU(3)\times SU(2)\times U(1) \equiv G_{\rm MSSM}$.  The gauge couplings $g_i$
of the remaining groups are given by $1/g_i^2 = 1/(g_{i,vis})^2 + 1/(g_{hid})^2$.
One linear combination of the gauginos is heavy due to the super-higgs mechanism.
The light gaugino, $\lambda_{\rm MSSM}$ is the linear combination
\begin{equation}
\frac{g_{vis}}{\sqrt{g_{vis}^2 + g_{hid}^2}}\lambda_{hid} -
           \frac{g_{hid}}{\sqrt{g_{vis}^2 + g_{hid}^2}}\lambda_{vis}.
\end{equation}
The soft gaugino masses for the three MSSM gauge groups are
\begin{equation}
m_i = \frac{\alpha_{i}}{\alpha_{hid}} m_{1/2}^{hid}
        = \frac{\alpha_{i}}{4\pi}\frac{F}{M}
\label{gauginomass}
\end{equation}
which is identical to that expected from canonical GMSB.

As discussed, the scalar masses above the scale $v$ are negligible. At the threshold $v$
small scalar masses are generated from integrating out the massive fields $L,\overline L$
and the heavy gaugino. The dominant contribution arises from a one-loop diagram with
gauginos and gives $\sim (\alpha/4\pi) (m_{1/2}^{hid})^2$. This
is smaller than the weak scale scalar masses generated from the renormalization group
by a loop factor.
We discuss theoretical issues regarding the threshold corrections to the scalar
masses in more detail in the following section. 

We see from Eq.~(\ref{gauginomass}) that gaugino masses appear to unify at the GUT 
scale. On the other hand, if one runs the ``observed'' scalar masses to higher scales
one would discover that they all vanish at an intermediate scale $v$.

\section{Scalar masses from threshold corrections}

In this section we give a more careful treatment of the calculation of the MSSM
scalar masses in order to show that they are really suppressed at the scale $v$.
We begin by integrating out the messenger fields $T+\overline T$ and write an
effective Lagrangian valid below the scale $M$. At one-loop
we get the gaugino mass
\beq
{\alpha_{hid}\over 4 \pi}\ \int d^2\theta \log(X)\ W_\alpha W^\alpha \longrightarrow 
       \ {\alpha_{hid}\over 4 \pi}\ {F \over M} \lambda \lambda\  \ .
\label{gaugkin}
\eeq
This is the only $d^2\theta$ term which can be generated by the non-renormalization
theorem. However, the diagram also generates higher-dimensional K\"ahler terms such as
\beq
{\alpha_{hid}\over 4 \pi}\ \int d^4\theta {X^\dagger X \over M^4}
     W_\alpha D^2 W^\alpha \ \longrightarrow
     \ {\alpha_{hid}\over 4 \pi}\ {F^\dagger F \over M^4} \lambda\! \dsl \lambda\ ,
\label{blob}
\eeq
a SUSY violating renormalization of the gaugino kinetic terms.
This is the leading operator, others are
suppressed by more powers of $M$ and decouple rapidly at lower energies.

\begin{figure}[htb]
\vskip 0.0truein
\centerline{\epsfysize=4in
{\epsffile{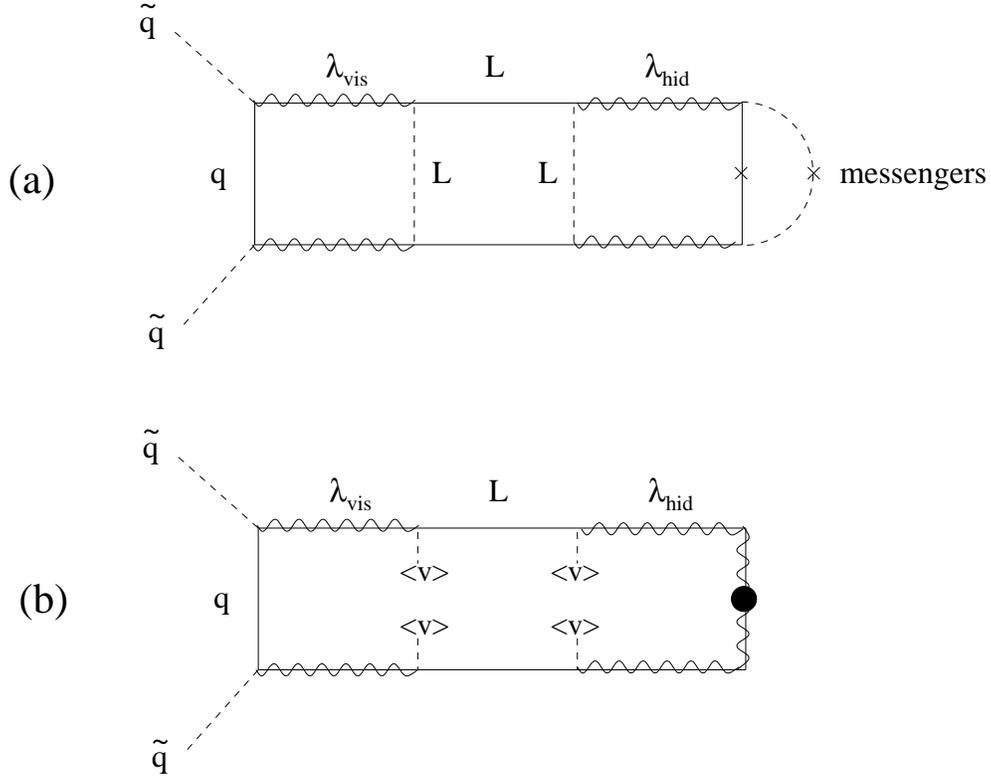}}}
\vskip 0.0truein
\caption[]{\it Loop contributions to the MSSM soft scalar masses, (a) above the scale $v$,
and (b) below that scale. The black dot represents an insertion of the SUSY violating
gaugino wave function operator, Eq.~(\ref{blob}).}
\end{figure}

The link scalars first couple to supersymmetry breaking at two loops and obtain the
usual gauge-mediated scalar mass $\sim (\alpha/ 4 \pi)^2 F^2/M^2$. The leading
diagram contributing to MSSM scalar masses at this scale is the four-loop diagram shown
in Figure 2(a). For loop momenta which are large compared to $v$ this diagram
yields $\sim (\alpha_{vis}/ 4 \pi)^2 (\alpha_{hid}/ 4 \pi)^2 F^2/M^2$. Note that
in the limit of either one of the two gauge couplings becoming large ($\sim 4\pi$),
this scalar mass contribution is important and the model does not reproduce the
gaugino mediation spectrum.  

Thus, above the scale $v$ MSSM scalar masses are suppressed by two loop factors
relative to the square of the gaugino masses. One might worry that after breaking
of the gauge groups to the diagonal much larger contributions to the scalar masses
may arise from the diagram in Figure 2(b).
Because of the large number of
propagators the diagram is UV-finite, it is dominated by loop momenta of order $v$ where we
can estimate its contribution to be $\sim v^2/M^2\ (\alpha_{hid} \alpha_{vis}/ 16 \pi^2)\ 
F^2/M^2$. Thus this contribution is
suppressed by the separation of scales between the messenger mass and the link
vevs. In the opposite limit, $v \gsim M$, the suppression disappears and the
diagram reduces gives the usual two-loop scalar masses of canonical GMSB. This
should be no surprise because for $v \gg M$ the breaking to the diagonal group
occurs first; below the scale $v$ the model {\it is} just canonical gauge mediation.

Finally, there is one more contribution to MSSM scalar masses from loop
momenta near $v$. It come from the same diagram in Figure 2(b). but with two
gaugino mass insertions. This gives $\sim (\alpha_{vis}/ 4 \pi) (\alpha_{hid}/ 4\pi)^2\ 
F^2/M^2$ which smoothly matches onto the contribution in the MSSM below $v$.

In summary, we find that as long as the scales $M$ and $v$ are well separated our
model reproduces the gaugino mediation spectrum. The MSSM scalar masses at the high scale
are suppressed relative to gaugino masses by the smaller of a
loop factor or the ratio $(v/M)^2$.

\section{Gauge coupling unification}

We now turn to the issue of gauge coupling unification. So far, we
have only dealt with energies below the unification scale. Therefore,
we assumed that $G_{vis}=SU(3)\times SU(2)\times U(1)$.
If coupling unification is not a numerical accident, then at the
unification scale $G_{vis}$ must be unified. For simplicity, we
will take $G_{vis}=SU(5)$ at that scale. The product group
structure turns out not to affect the unification scale.

Below the scale $v$, the gauge couplings evolve according to ordinary
MSSM equations. At $v$, $G_{vis}$ and $G_{hid}$ are no longer broken to their
diagonal subgroup and the couplings evolve independently.
Thus, the Standard Model gauge couplings need to be matched onto
the new theory. At tree level we have
\begin{equation}
   \frac{1}{{g}_i^2}=\frac{1}{g_{i,vis}^2}+\frac{1}{g_{hid}^2},
\end{equation}
where $g_i$ indicate MSSM couplings just below $v$, and $g_{i,vis}$
corresponding $G_{vis}$ couplings just above $v$. All $1/g_i^2$
are shifted by the same amount, $1/g_{hid}^2$. Therefore,
the differences between the gauge couplings remain unchanged across the
threshold (this fact is also exploited in \cite{weiner}).

Above $v$, the link fields $L,\overline{L}$ also contribute to
the running in addition to the gauge bosons. The link fields
form complete multiplets of $SU(5)_{vis}$, so they do not alter the
relative running of the gauge couplings. Since the differences
of couplings are not changed across the threshold $v$ and continue
to evolve at the MSSM rate, the unification scale for $G_{vis}$
is unchanged and equals to $2\cdot 10^{16}~{\rm GeV}$.

\section{Conclusions}

In this letter we present a simple, calculable, renormalizable four-dimensional
theory of SUSY breaking and mediation which reproduces the superpartner
spectrum of gaugino mediation.
There are no branes, troublesome moduli or dynamical assumptions.
The superpartner masses are generated from field theory dynamics. They are
flavor preserving and
insensitive to uncertainties stemming from our incomplete understanding
of quantum gravity as long as the messenger scale $M \lsim 10^{-4} M_{Planck}$.\
As in higher dimensional gaugino mediation
the requirements on the SUSY breaking sector are much less stringent than
in canonical GMSB, we only require that the gauginos of the hidden sector 
unsuppressed masses. 

The superpartner spectrum is determined by the fact that the
MSSM scalar masses vanish at the scale $v \ll M$. Weak scale
scalar masses are predicted from the RG evolution \cite{ginomsb,baer},
and become similar to no-scale \cite{no-scale,YY} masses
when $v$ approaches $M_{GUT}$. 
Roughly, one finds that the masses of colored fields are largest while
fields with only hypercharge couplings are smallest. Therefore the right-handed
sleptons and the Bino are light, with the stau being the lightest MSSM superpartner.
This spectrum is preferred by fits to the recent measurement of the muon anomalous magnetic
moment \cite{gminus2}.
In contrast with minimal gaugino mediation we find that
the gravitino is the LSP as in GMSB.

\section{Acknowledgements}

We thank Csaba Csaki and Graham Kribs for useful discussions.
DK and WS thank the summer visitor program of the Theory Department at Fermilab for
their hospitality. H.-C. Cheng is supported by the Robert R. McCormick Fellowship. We
are also supported by the DOE grants DE-FG02-90ER-40560 (HCC, DEK),
W-31-109-ENG-38 (DEK), DE-AC02-76CH03000 (MS), DE-FC02-94ER40818 (WS).


\end{document}